\documentclass[12pt]{article}

\usepackage{graphicx}
\title{Statefinder diagnosis of nearly flat and thawing non-minimal quintessence}
\author{A. Shojai and F. Shojai\\
Department of Physics, University of Tehran, Tehran, Iran.} 
\date{}
\begin{document}
\maketitle
\begin{abstract}
Non-minimally coupled scalar field models of dark energy are equivalent to an interacting quintessence in the Einstein's frame. Considering two special important choices of the potential of the scalar field, i.e. nearly flat and thawing potential, one has an analytical expression for the equation of state parameter as a function of the density parameter of the scalar field for any choice. Here we investigate the non-minimal quintessence model by applying the method of statefinder diagnosis to it and plotting the evolutionary trajectories of the statefinder parameters.
\end{abstract}
\section{Introduction}
According to the supernova observations \cite{1feng}, our universe has an accelerating expansion at present epoch \cite{1track,pad}. Accepting general relativity theory as the underlying theory for describing the dynamics of the universe, there is a missing energy-momentum component of the universe, dubbed "dark energy" \cite{3thawing}, providing the present acceleration. There are several candidate for such a dark energy, cosmological constant, chaplyian gas, quintessence and so on (See Ref \cite{3thawing} and references therein).

Some cosmological models involve an interaction between dark matter and dark energy \cite{7}. This removes the need for a fine-tuned cosmological constant to get the ratio of dark matter to dark energy of order unity nowadays. This is widely known as coincidence problem \cite{6}.

A particular class of these models is Brans-Dicke (BD) type scalar field model. 
One can deal with BD scalar field model in two frames,"Jordan frame" and "Einstein frame".  In Jordan frame, the the scalar field does not appear in the action of matter fields but couples non-minimally to gravity. It is possible to make a particular conformal transformation, converting the theory to a  scalar-tensor model in which the scalar field couples conformally to matter and minimally to gravity. Describing the scalar field as quintessence, hereafter we shall refer to this model as conformally coupled quintessence model (CCQ). There has been a lot of discussion about the choice of one of these frames as the physical frame \cite{7track,8track}. Here we pay our attention to the Einstein frame. This frame is a suitable framework to make the model consistent with the solar system constraint and further observational constraints from Big-Bang nucleosynthesis and inflation \cite{8thawing}. 

Following Caldwell and Linder \cite{11thawing}, the scalar field models of quintessence can be divided into two categories, called "freezing" and "thawing" models. In the former, the equation of state parameter has a decreasing behavior but in the latter, it has a value near $-1$ initially, and then increases with time to less negative values.

In \cite{19thawing}, it is shown that non-interacting thawing quintessence model with nearly flat potentials (the potentials satisfying the slow-roll conditions: $\left(\frac{1}{V}\frac{dV}{d\varphi}\right)^2\ll1, \frac{1}{V}\frac{d^2V}{d\varphi^2}\ll1$)  
provides a natural way to produce a value of $\omega$ near $-1$ today. Generalizing this to CCQ model with nearly flat potentials \cite{18thawing}, shows that there exists
a universal behavior for $\omega$ which is different from thawing behavior initially. In a recent paper \cite{thawing}, some conditions on the potential of the scalar field are derived which is different from the slow-roll conditions initially and lead to the thawing behavior for all times.

On the other hand as demonstrated in Refs.\cite{state}, it is possible to discriminate different models of dark energy from each other by some parameters, called the statefinder parameters $(r,s)$, firstly proposed in \cite{sahni2003} and \cite{alam}. The first of these, $r$, is the jerk parameter and the other is a function of jerk and the decelerating parameters. The statefinder pair depends on the metric of space-time and is constructed using the second and third derivatives of the scale factor as:
\begin{equation}
r=\frac{\dot{\ddot{a}}}{a H^3}
\end{equation}  
\begin{equation}\label{s}
s=\frac{r-1}{3(q-1/2)} 
\end{equation}
where $q=-\ddot{a}/{\dot{a}}^2$ is the deceleration parameter. 

In this paper we will discuss further the nearly flat and thawing CCQ models of dark energy by statefinder diagnostic. In the next section, we will start with BD interacting dark energy model. According to \cite{18thawing} and \cite{thawing}, application of  some conditions on the 
potential, nearly flat and thawing conditions, leads to an analytical expression for $\omega$ in any case. Then, we apply the statefinder diagnostic to CCQ model in section 3. In the last section we will give some conclusions. Throughout this work we have chosen the units $8\pi G=c=1$. 
\section{The model}
The general action of BD scalar tensor theory in the Jordan frame is:
\begin{equation}
S_{J}=\int d^{4}x
\sqrt{-\tilde{g}}\left[\Phi\tilde{R}-\frac{\tilde{\omega}}{\Phi}\Phi^{,\mu}\Phi_{,\mu}-2U(\Phi)+\mathcal{L}_{m}(\tilde{g}_{\mu\nu})\right].
\end{equation}
where $\tilde{R}$ is the Ricci scalar of the metric $\tilde{g}_{\mu\nu}$, $U(\Phi)$ is the potential of the scalar field, $\tilde\omega$ is the BD coupling constant and $\mathcal{L}_{m}$ is the matter Lagrangian. Under the conformal transformation
$g_{\mu\nu}=e^{\zeta\varphi}\tilde{g}_{\mu\nu}$
 , in which $\ln \Phi=\zeta\varphi$ and $\zeta=\sqrt{\frac{2}{3+2\tilde\omega}}$, one arrives at the action of BD theory in  the Einstein frame:
\begin{equation}
S_{E}=\int d^{4}x
\sqrt{-g}[R-\frac{1}{2}(\nabla\varphi)^{2}
-V(\varphi)+\mathcal{L}_{m}(e^{-\zeta\varphi}g_{\mu\nu})].
\end{equation}
where $V(\varphi)=e^{-2\zeta\varphi}U(\Phi(\varphi))$. 

Now, consider a spatially flat FRW universe occupied by the pressureless matter. The cosmological equations of motion are:
\begin{equation}\label{friedmann}
H^{2}=\frac{1}{3}(\rho_{\varphi}+\rho_{m})
\end{equation}
\begin{equation}\label{friedmann1}
\dot{H}=-\frac{1}{2}\left[\rho_{m}+\rho_{\varphi}+p_{\varphi}\right]
\end{equation}
and the scalar field evolution is governed by the following equation of motion:
\begin{equation}\label{scalar field}
\ddot{\varphi}+3H\dot{\varphi}+V_{\varphi}=\sqrt{\frac{2}{3}}\beta\rho_{m}
\end{equation}
where $\rho_{m}$, $\rho_{\varphi}=\frac{1}{2}\dot{\varphi}^{2}+V(\varphi)$ and
$p_{\varphi}=\frac{1}{2}\dot{\varphi}^{2}-V(\varphi)$ denote the energy density of cosmic fluid, the energy density and pressure density of the scalar field in the Einstein frame respectively and $\beta=\sqrt{\frac{3}{8}}\zeta$. From equation (\ref{scalar field}), one can easily see that the energy density of the scalar field satisfies the following conservation law:
\begin{equation}\label{rhophidot}
\dot{\rho}_{\varphi}+3H(1+\omega)\rho_{\varphi}=\sqrt{\frac{2}{3}}\beta\dot{\varphi}\rho_{m}
\end{equation}
in which $\omega=p_{\varphi}/\rho_{\varphi}$ is the equation of state parameter for the scalar field. Combining the above equations, the continuity equation for the cosmic fluid can be derived as:
\begin{equation}\label{rhomdot}
\dot{\rho}_{m}+3H\rho_{m}=-\sqrt{\frac{2}{3}}\beta\dot{\varphi}\rho_{m}
\end{equation}
By integrating equation (\ref{rhomdot}), one obtains:
\begin{equation}\label{rhom}
\rho_{m}(t)=\rho_{0m}\left (\frac{a}{a_0}\right )^{-3}e^{-\frac{(\varphi-\varphi _0)}
{\sqrt 6}}
\end{equation}
in which $\rho_{0m},a_{0}$ and $\varphi_0$ are the current values of the matter density, the scale factor and the scalar field respectively. Therefore the usual dependence of the non-relativistic matter upon the scale factor is modified by an exponential factor due to the interaction with the scalar field.

Taking the time derivative of $V(\varphi)=\frac{(1-\omega)}{2}\rho_{\varphi}$ and using the continuity equation (\ref{rhophidot}), one obtains \cite{tracker}:
\begin{equation}\label{V1}
\omega'=-3(1-\omega^2)\left[1-\frac{\sqrt{\Omega_\varphi}}{\sqrt{3(1+\omega)}}\left(-\frac{V_{\varphi}}{V}+\frac{1}{\sqrt 6} \frac{1-\Omega_\varphi}{\Omega_\varphi}\right)\right]
\end{equation}
where $\Omega_{m}$ and $\Omega_{\varphi}$ are the density parameters of matter and the scalar field respectively ($\Omega_{i}=\rho_{i}/3 H^2)$ and prime denotes derivative with respect to $\ln a$.. Ignoring the interaction of the scalar field and the cosmic fluid, this equation reduces to the corresponding equation of Steinhardt et al. \cite{13track} for a non interacting scalar field.

Following the same method as refs \cite{19thawing,18thawing},  using the relation $\dot{\phi}^2=(1+\omega)\rho_\phi$, the continuity equation (\ref{rhophidot}) reads as \cite{thawing}:
\begin{equation}\label{z}
\Omega'_\phi=3(1-\Omega_\phi)\left(-\omega\Omega_\phi+\frac{1}{6}\sqrt{2(1+\omega)\Omega_\phi}\right)
\end{equation}
which is a useful relation to derive the redshift dependence of the cosmological quantities in the next section.

Moreover by using equation (\ref{rhom}), one can express equations (\ref{friedmann}) and (\ref{friedmann1}) in the following form, convenient for the construction method proposed by \cite{sahni}:
\begin{equation}
\frac{1}{9{H_0}^2}V(x)=\frac{H^2}{{H_0}^2}-\frac{x}{6{H_0}^2}\frac{dH^2}{dx}-\frac{1}{2}\Omega_{0m}e^{-\frac{1}{\sqrt 6}(\varphi-\varphi _0)}x^3
\end{equation}
\begin{equation}\label{res}
\frac{1}{9{H_0}^2}\left({\frac{d\varphi}{dx}}\right)^2=\frac{2}{3x{H_0}^2}
\frac{d\ln H}{dx}- \frac{\Omega_{0m}e^{-\frac{1}{\sqrt 6}(\varphi-\varphi _0)}x}{{H_0}^2}
\end{equation}
where $x=1+z$ in which $z=a_0/a-1$ is the redshift parameter and $a_0$ is the present scale factor of the universe. These are two coupled equations allowing one to reconstruct the potential $V(z)$ and the scalar field $\varphi(z)$ knowing $\Omega_{0m}$, $\varphi _0$ and also using the observed $H(z)$ from the luminosity distance.

Since the left hand side of equation (\ref{res}) is non-negative,  the dynamical expansion of the universe is restricted by the following inequality:
\begin{equation}\label{ineq}
\frac{d{H}^2}{dz}\geq 3H_{0}^2\Omega_{0m}(1+z)^{2}e^{-\frac{1}{\sqrt 6}(\varphi-\varphi _0)}
\end{equation}
This is the weak energy condition for BD cosmology in the Einstein's frame. We see that although it is not possible to express the Hubble parameter as an explicit function of the cosmological redshift in this model, there is a necessary condition on the $H(z)$ in this case. Ignoring the interaction and thus the exponential factor, inequality (\ref{ineq}) reduces to what is derived by Sahni and Starobinsky in \cite{1track} for non interacting case.  
\section{Statefinder diagnostic of CCQ model of dark energy} 
According to the previous section, using BD theory of gravity in the Einstein's frame leads to a scalar-tensor theory in which there is a particular interaction between the non-relativistic matter and the scalar field. It is a well known fact that the coupled scalar field model with $-1\leq\omega\leq-1/3$ (called usually interacting quintessence) has the capability of explaining the current cosmic acceleration \cite{2new}. However to make these models acceptable, some limitations on the form of $V$  has to be set, depending on the form of the interaction term \cite{13track,11track,12track}.

In \cite{thawing}, the necessary conditions for the existence of thawing behavior for CCQ model is found. Assuming thawing behavior, the following relation, called thawing condition, should be satisfied \cite{thawing}:
\begin{equation}\label{co}       
\lambda\equiv-\frac{V_{,\varphi}}{V}\simeq\lambda_0-\frac{1}{\sqrt6}\frac{\Omega_m}{\Omega_{\varphi}}
\end{equation}
which shows that it is necessary that $\lambda$ increases with time when $\varphi$ and $\Omega_{\varphi}$ are increasing functions. Dividing equation (\ref{V1}) by equation (\ref{z}), one arrives at a differential equation for $\omega$ as a function of $\Omega_{\varphi}$ \cite{thawing}. Replacing $\lambda$ with expression (\ref{co}) and retaining terms to the lowest order in $1+\omega$ ($\omega$ is near $-1$), the differential equation of $\omega$ is exactly solvable. The resulting analytical expression expression for the state parameter is as follows \cite{thawing}:
\begin{equation}\label{ome}
1+\omega=\left(\frac{1-\Omega_{\varphi}}{\Omega_{\varphi}}\right)^{2A}\left[\chi_0+\frac{2\lambda_0\Omega_{\varphi}^{1/2+A}}{\sqrt3(1+2A)}{}_2F_1(\frac{1}{2}+A,1+A,\frac{3}{2}+A,\Omega_\varphi)\right]^2
\end{equation}
where $\lambda_0$ is a positive constant, $\chi_0$ is an integration constant, $A=1
+\beta\sqrt{\frac{2}{27}}\lambda_0$ and ${}_2F_1$ is the Gauss Hypergeometric function.

In \cite{thawing} it is shown that in the case of CCQ model, $\lambda$ neither is a small value nor is  a constant. Therefore the nearly flat potentials  do not lead to the thawing behavior.
However recently, study of the behavior of the equation of state parameter for nearly flat potentials  has attracted a lot of attention \cite{19thawing,18thawing}. 
The authors of \cite{18thawing} have shown that the equation of state parameter  firstly increases with time and then approaches asymptotically to a value near to $-1$. As mentioned in \cite{18thawing}, assuming the slow-roll conditions for the potential, one can show that $\left|\frac{\lambda'}{\lambda}\right|\ll1$ which ensures that $\lambda$ is approximately constant up to now, i.e. 
\begin{equation}\label{con}
\lambda\simeq\lambda_0=\left . -\frac{V_{,\varphi}}{V}\right |_{\varphi_0}
\end{equation}
where $\lambda_0$ is a small constant evaluated at the initial value of $\varphi_0$. Combining equations (\ref{V1}) and (\ref{z}) and making two assumptions, the first one is that $\omega$ is near $-1$ and the second is that the condition (\ref{con}) is satisfied, these yield again to a differential equation for $\omega$ as a function of $\Omega_{\varphi}$ \cite{18thawing} which gives the following analytical expression for the equation of state parameter:
\begin{equation}\label{omeg}
1+\omega=[\frac{\lambda_0}{\sqrt{3\Omega_\varphi}}-\left(\frac{1}{\Omega_\varphi}-1\right)[(\frac{\lambda_0}{2\sqrt3}-\frac{\sqrt2 \beta}{3})\ln\left(\frac{1+\sqrt{\Omega_\varphi}}{1-\sqrt{\Omega_\varphi}}\right)-\alpha]]^2
\end{equation}
where $\alpha=-\frac{\lambda_0}{\sqrt3}\frac{2\sqrt\Omega_i\Omega_i}{1-\Omega_i}-\frac{2\sqrt2\beta}{3}\Omega_i$ in which $\Omega_i$ is some small initial value of $\Omega_\varphi$ such that $\omega_i=-1$.  

From equations (\ref{friedmann}) and (\ref{friedmann1}), it is straightforward to show that the deceleration parameter takes the following form as:
\begin{equation}\label{q}
 q=\frac{1}{2}(1+3\omega\Omega_{\varphi})
\end{equation}
After differentiating equation (\ref{friedmann1}), using relations (\ref{rhophidot}) and (\ref{rhomdot}), one finds:
\begin{equation}\label{r}
r=1+\frac{9}{2}\omega(1+\omega)\Omega_{\varphi}-\frac{3}{2}\omega'\Omega_{\varphi}-\frac{3}{2\sqrt{2}}\omega\sqrt{1+\omega}\sqrt{\Omega_{\varphi}}(1-\Omega_{\varphi})
\end{equation}
Inserting (\ref{q}) and (\ref{r}) in (\ref{s}) gives:
\begin{equation}\label{rr}
s=1+\omega-\frac{\omega'}{3\omega}-\frac{1}{3\sqrt{2}}\frac{1-\Omega_\varphi}{\sqrt{\Omega_\varphi}}\sqrt{1+\omega}
\end{equation}
which explicitly depends on $\Omega_{\varphi}$ in contrast to the non-interacting quintessence.

Let us now have a detailed look at how the statefinder parameters behave for thawing and nearly flat CCQ model of dark energy. As we have seen before, in these cases, it is possible to derive an analytical expression for the statefinder pairs as a function of the density parameter of the scalar field, $\Omega_{\phi}$, without considering a special form for the potential. To do this, one can use the relation (\ref{V1}), the thawing condition (\ref{co}) and the analytical expression of $\omega$, equation(\ref{ome}), for the case of thawing CCQ model and the corresponding relations (\ref{V1}), (\ref{con}) and (\ref{omeg})  in the case of nearly flat CCQ model.  

\begin{figure}
\begin{center}
\includegraphics[scale=0.65]{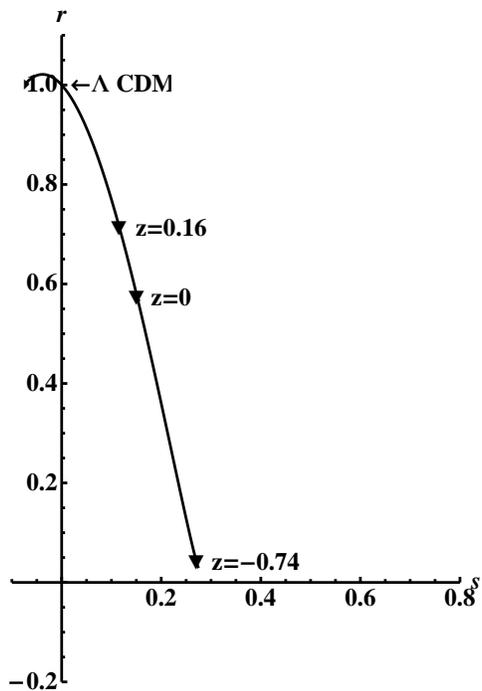} 
\end{center}
\caption{Plot of $r$ versus $s$ for thawing model.}
\label{thaw1}
\end{figure}

\begin{figure}
\begin{center}
\includegraphics[scale=0.65]{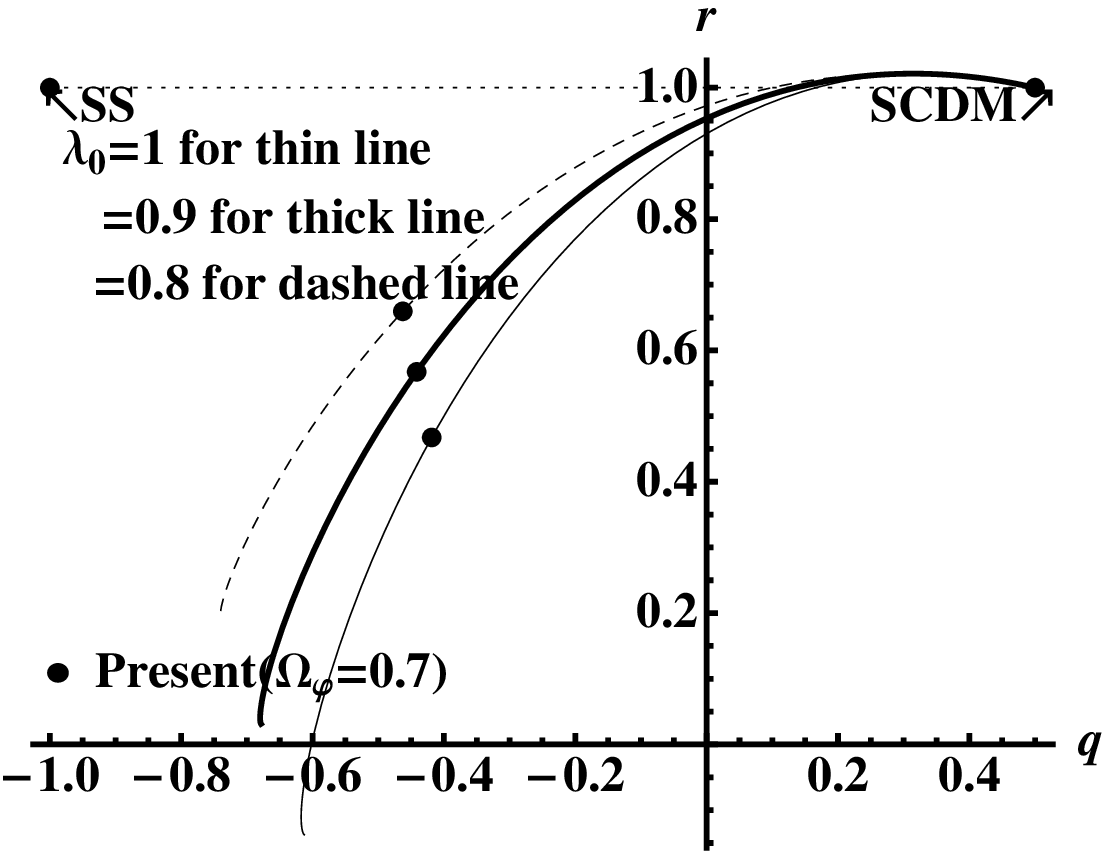} 
\end{center}
\caption{Plot of $r$ versus $q$ for thawing model.}
\label{thaw2}
\end{figure}

The time evolution of the statefinder pairs $(r,s)$ for thawing CCQ model has been shown in Figure \ref{thaw1} with $\lambda_0=0.9$. This value of $\lambda_0$ is chosen such that $\omega$ has a value near $-1$ today. Also the constant $\chi_0$ is chosen such that the initial condition $\omega=-1$ holds for $\Omega_{\phi}=0.001$ and $\beta$ has been set equal to $0.5$. $\Lambda CDM$ ($\Lambda$-cold dark matter) universe corresponds to the fixed point $(1,0)$. We see that the evolution of trajectories of statefinders pairs pass from
the point $(r,s)\simeq(1,-0.07)$ in the past when $z\simeq4.62$ and $\Omega_\phi\simeq0.01$. And then after passing $\Lambda\textit{CDM}$ fixed point, $r$ decreases whereas $s$ increases to the point $(r,s)\simeq(0.04,0.27)$ at $z\simeq-0.74$ $(\Omega_\phi\simeq0.99)$ in the future. The location of today's point is $(r,s)\simeq(0.57,0.15)$ when $\Omega_\phi\simeq0.7$. This shows the present 'distance' of thawing CCQ model of dark energy from $\Lambda\textit{CDM}$ model. The time evolution of the pairs $(r,q)$ is indicated by Figure \ref{thaw2}. The dashed, thick and thin curves have $\lambda_0=0.8,0.9,1$ respectively. We see that both $\Lambda\textit{CDM}$ and thawing CCQ models start evolving from the same point, $(r,q)=(1,0.5)$ which corresponds to SCDM (standard cold dark matter) universe. In $\Lambda\textit{CDM}$ scenario, the evolution is along a horizontal line ends at SS (steady-state) fixed point, $(r,q)=(1,-1)$, corresponding to de Sitter expansion. However in our model, $q$ has a decreasing behavior whereas the value of $r$ first increases and then monotonically decreases. The trajectory goes to $(r,q)\simeq(0.04,-0.68)$ at $z\simeq-0.7$ in the future. 

In Figures \ref{thaw4} and \ref{thaw5}, we have shown the deceleration parameter and the equation of state parameter as functions of redshift parameter. These figures have been plotted numerically using the relations (\ref{z}), (\ref{ome}) and (\ref{q}). 

\begin{figure}
\begin{center}
\includegraphics[scale=0.65]{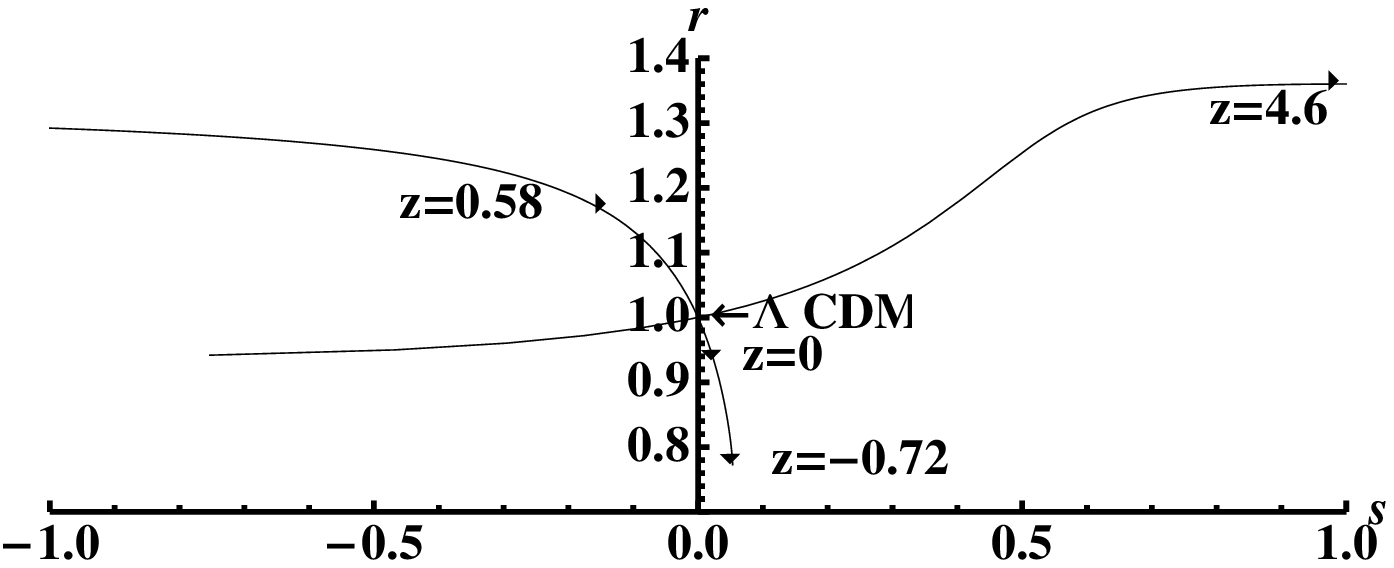} 
\end{center}
\caption{Plot of $r$ versus $s$ for nearly-flat model.}
\label{flat1}
\end{figure}

In Figure \ref{flat1} we have shown time evolution of the statefinder 
pairs for nearly flat CCQ model of dark energy with $\lambda_0=0.4$. This small value of $\lambda_0$ ensures that the variation of the potential during the evolution of the universe is very small. Assuming that for an initial value of $\Omega_\phi$, say $\Omega_i$, $\Omega_i=0.001$, we have $\omega=-1$. This determines the constant $\alpha$ in the relation
(\ref{omeg}) setting $\beta=0.5$ here. The $(r,s)$ trajectory has two branches. We find that the first branch comes asymptotically from $r\simeq 1$, $s\rightarrow-\infty$
goes to $r\simeq 1.33$, $s\rightarrow+\infty$ for a change of $\Omega_\phi$ in the interval $[0.001,0.17]$. Another branch comes 
along $r\simeq 1.33$ asymptote, passes from $\Lambda\textit{CDM}$ point and then goes
to $(r,s)\simeq(0.78,0.05)$ at $z\simeq-0.7$ ($\Omega_\phi\simeq0.99$) in the future. The divergent behavior and discontinuity of $s$ occurs at $\Omega_\phi\simeq 0.17$ and is due to vanishing $\omega$ at  this point. Along the first branch, both $r$ and $s$ are increasing, however for the other branch $r$ monotonically decreases whereas $s$ increases. The location of today's point is $(r,s)\simeq(0.94,0.02)$ when $\Omega_\phi\simeq0.7$. This shows that this model has less distance from $\Lambda\textit{CDM}$ model in comparison to thawing model. This result is satisfied in both models independent of the allowed chosen value of $\lambda_0$.

\begin{figure}
\begin{center}
\includegraphics[scale=0.65]{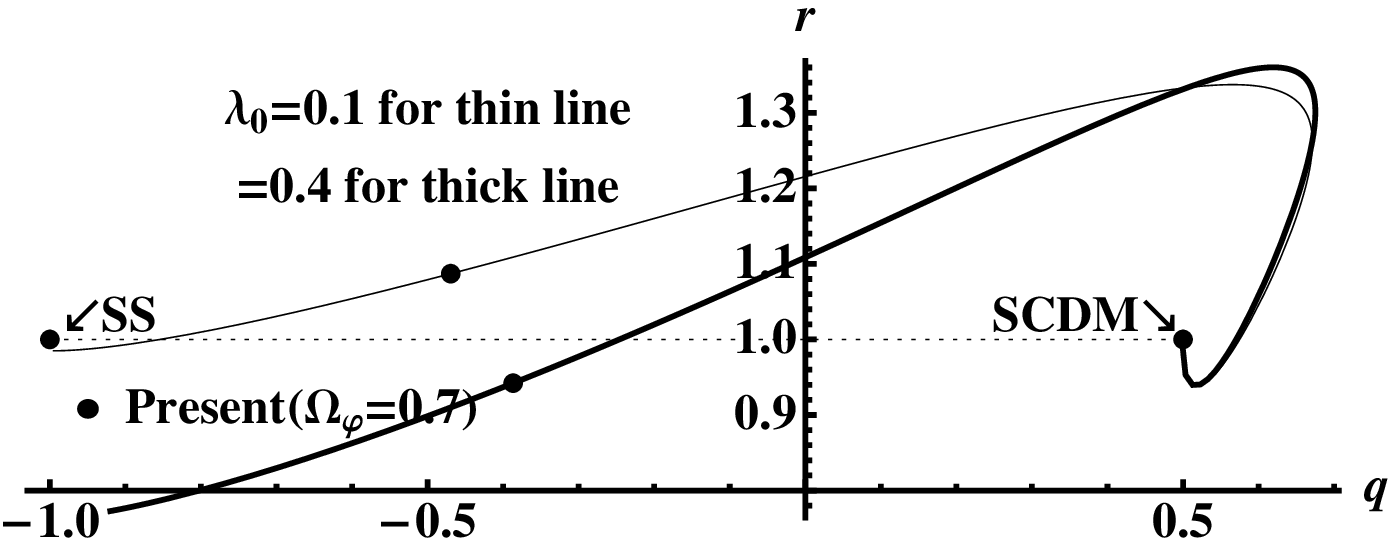} 
\end{center}
\caption{Plot of $r$ versus $q$ for nearly-flat model.}
\label{flat2}
\end{figure}

The evolutionary track in $(r,q)$ plane is shown in Figure \ref{flat2}. The tick and thin curves are corresponding to $\lambda_0=0.4$ and $0.1$ respectively. It is started from SCDM point as the same as the $\Lambda\textit{CDM}$ 
evolutionary path, but first $r$ decreases and $q$ increases slightly, then after passing a period in which $r$ and $q$
have increasing behavior, those decreases monotonically. This period corresponds to very high redshift. The trajectory goes to $(r,q)\simeq(0.78,-0.9)$ at $z\simeq-0.71$ ($\Omega_\phi\simeq0.99$) in the future. At the end, we present the plot of $q$ and $\omega$ with respect $z$.
\section{conclusion} 
\begin{figure}
\begin{center}
\includegraphics[scale=0.65]{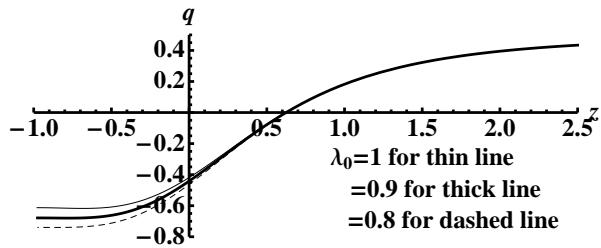}
\end{center} 
\caption{Plot of $q$ versus redshift for thawing model.}
\label{thaw4}
\end{figure}

\begin{figure}
\begin{center}
\includegraphics[scale=0.65]{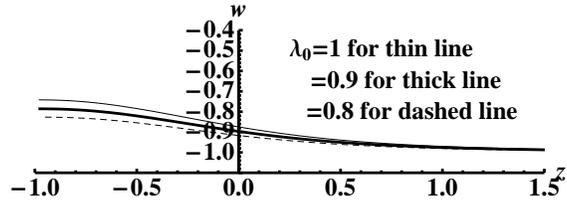}
\end{center} 
\caption{Plot of $\omega$ versus redshift for thawing model.}
\label{thaw5}
\end{figure}

\begin{figure}
\begin{center}
\includegraphics[scale=0.65]{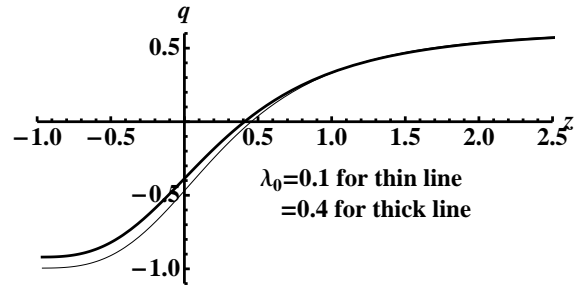} 
\end{center}
\caption{Plot of $q$ versus redshift for nearly-flat model.}
\label{flat4}
\end{figure}

\begin{figure}
\begin{center}
\includegraphics[scale=0.65]{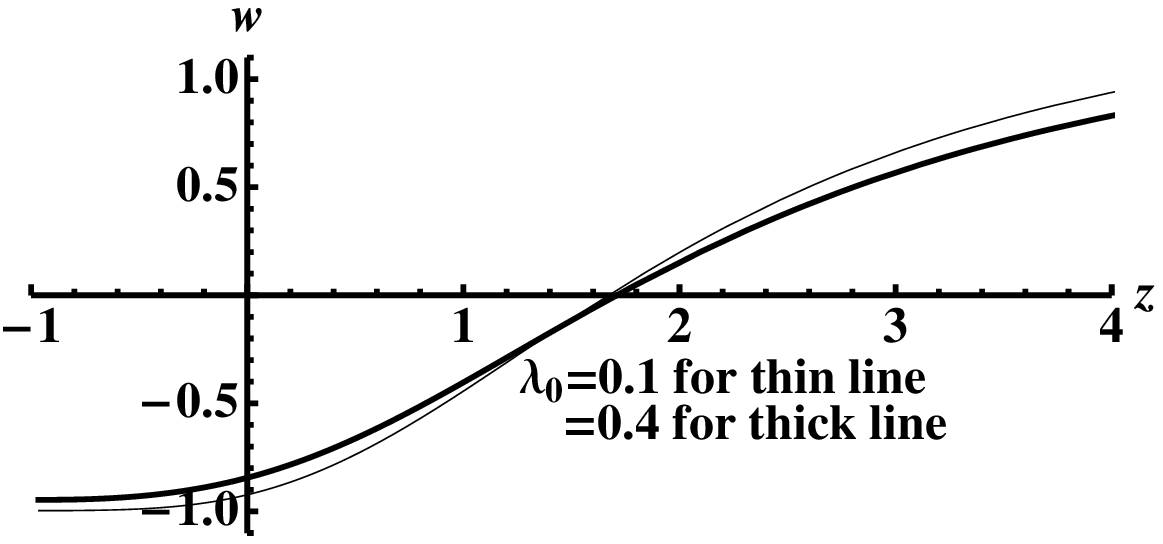} 
\end{center}
\caption{Plot of $\omega$ versus redshift for nearly-flat model.}
\label{flat5}
\end{figure}

In this paper, we have applied the statefinder diagnostic to the CCQ model of dark energy. And we have plotted the trajectories in the $(r,s)$, $(r,q)$, $(q,z)$ and $(\omega,z)$ for nearly flat and thawing potentials. As it is apparent from Figures \ref{thaw4} and \ref{flat4}, the deceleration parameter decreases 
monotonically with redshift for both the nearly flat and thawing
potentials, however it goes to more negative values at future in the case of nearly flat model. Moreover nearly flat potentials force the equation of state parameter changes in a wide range
with redshift whereas it remains near $-1$ for thawing model, as one expected. This is exactly what one finds from Figures \ref{thaw5} and \ref{flat5}. Also the value of $r$ decreases with redshift steadily for both kinds of potentials as well as non-interacting quintessence model \cite{sahni2003}.

\textbf{Acknowledgment} This work is partly supported by a grant from university of Tehran and partly by a grant from center of excellence of department of physics on the structure of matter.

\end{document}